\documentclass[twocolumn,showpacs,pra,aps,superscriptaddress,amsmath,amssymb,floatfix,]
{revtex4}
\usepackage{graphicx}
\begin{document}
\title{Quantum-wave evolution in a step potential barrier}
\author{Jorge Villavicencio}
\email{villavics@uabc.mx}
\affiliation{Instituto de F\'{\i}sica,
Universidad Nacional Aut\'onoma de M\'exico\\
Apartado Postal {20 364}, 01000 M\'exico, Distrito Federal, M\'exico}
\affiliation{Facultad de Ciencias,
Universidad Aut\'onoma de Baja California\\
Apartado Postal 1880, 22800 Ensenada, Baja California, M\'exico}
\author {Roberto Romo}
\email{romo@uabc.mx}
\affiliation{Facultad de Ciencias,
Universidad Aut\'onoma de Baja California\\
Apartado Postal 1880, 22800 Ensenada, Baja California, M\'exico}
\author {Sukey Sosa y Silva}
\email{sukeys@hotmail.com}
\affiliation{Facultad de Ciencias,
Universidad Aut\'onoma de Baja California\\
Apartado Postal 1880, 22800 Ensenada, Baja California, M\'exico}
\date{\today}
\begin{abstract}
By using an exact solution to the time-dependent 
Schr\"{o}dinger equation with a point source initial condition, we 
investigate both the time and spatial  dependence of quantum 
waves in a step potential barrier.
We find that for a source with energy below the barrier height, 
and for distances larger than the penetration length, the probability 
density exhibits a {\it forerunner} associated with a non-tunneling 
process, which propagates in space at exactly the semiclassical group 
velocity.
We show that the time of arrival of the maximum of the {\it forerunner} 
at a given fixed position inside the potential is exactly the traversal
time, $\tau$.
We also show that the spatial evolution of this transient pulse
exhibits an invariant behavior under a rescaling process. 
This analytic property is used to characterize the evolution of the 
{\it forerunner}, and to analyze the role played by the time of arrival, 
$3^{-1/2}\tau$, found recently by Muga and B\"{u}ttiker 
[Phys. Rev. A {\bf 62}, 023808 (2000)].
\end{abstract}

\maketitle

\section{INTRODUCTION}

Since the original proposal by Stevens\cite{Stevens} of 
tunneling monochromatic fronts, the problem has become controversial and
still open to investigations. The existence of such propagating fronts has
been supported in asymptotic analysis based on exact analytical solutions 
\cite{Moretti} to the time-dependent Schr\"{o}dinger equation, with point
source initial conditions. However, analytical and numerical studies based
on different boundary conditions and potentials \cite{todos}, have shown
evidence that it is not possible to identify such fronts, in the way 
suggested in Ref. \cite{Moretti}. 
Despite these clarifying works, the need of a direct and comprehensive study
of the original point source problem has been recently emphasized 
\cite{ButtThom,MugBu}. 

A recent study of Muga and B\"{u}ttiker \cite{MugBu} 
of quantum waves in a potential step, has shown the existence of a 
traveling transient {\it forerunner} 
for frequencies of the source below the barrier height, which in 
opaque barrier conditions ($x q_0\gg 1$) is dominated by over-the-barrier 
frequencies, implying that this transient structure is associated to a 
non-tunneling process. They also found that the time of arrival at 
a fixed position $x_f$ is unexpectedly given by ${3^{-1/2}}\tau$, where
$\tau$ is the {\it traversal time}, defined as $\tau=(x_f/v_{q_0})$, 
with $v_{q_0}=(\hbar q_0/m)$, and $q_0=[2m(V_0-E_0)]^{1/2}/\hbar$; the
parameters $V_0$ and $E_0$ correspond respectively to the step potential 
height and the initial energy of the source. We believe that in order to 
gain more insight about the evolution of quantum waves, the above investigation 
should be extended to study the role played by $\tau$ in the spatial-dependence
of the solution.
   
In this paper we investigate both the time and position dependence of quantum
waves in a step potential barrier by means of an exact solution
to the time-dependent Schr\"{o}dinger equation. 
Our analysis shows that the {\it forerunner} travels in 
space at exactly the semiclassical group velocity $v_{q_0}$, implying
that the traversal time $\tau$ is exactly the time of arrival of the 
maximum of the {\it forerunner} at a fixed point deep inside the potential 
step; clearly this result is different from the time scale of Ref. \cite{MugBu} 
by a factor of $3^{-1/2}$. Using a scaling property of the {\it forerunner} 
derived from the analytic solution, the exact interpretation of the time scales
$\tau$ and $3^{-1/2}\tau$ as times of arrival is clarified.
It is a surprising fact that $\tau$, a time scale traditionally regarded as one 
of the possible tunneling times, is here associated to a non-tunneling process.

This work is organized as follows. Section II deals with numerical examples, 
showing both spatial and time evolution of the probability density for 
initial waves with energies $E_0>V_0$, and $E_0<V_0$. Finally, the 
conclusions are presented in Section III.

\section{QUANTUM WAVE EVOLUTION}

In this section we shall explore the main features of quantum wave evolution. 
Here we use the exact analytical solutions \cite{Moretti,MugBu} (see also Appendix A) 
to the time-dependent Schr\"{o}dinger equation for a step potential barrier 
$V(x)=\Theta(x)V_0$, for a wave formed by a point source with a sharp onset,
\begin{equation}
\psi _{0}\left( x=0,t\right) =\left\{ 
\begin{array}{cc}
{\rm e}^{-i\omega _{0}t}, & \quad t>0, \\ 
0, & \quad t<0.
\end{array}
\right.   
\label{ci}
\end{equation}
We have defined $\omega_0=(E_0/\hbar)$, where $E_0$ corresponds to the initial
energy of the source.
It is worthwhile to point out that the source boundary condition given by 
Eq. (\ref{ci}) is not a standard one in quantum mechanics. For a physical 
interpretation of these type of boundary conditions see Ref. \cite{BEM01}.

The analytical solutions for the point source problem are given by,
\begin{equation}
\psi_>(x,t)={\rm e}^{-iVt}\left[ M(x,k_0,t)+M(x,-k_0,t)\right] ,\quad
\omega_0>V,
\label{propagation}
\end{equation}
and,
\begin{equation}
\psi_<(x,t)={\rm e}^{-iVt}\left[ M(x,iq_0,t)+M(x,-iq_0,t)\right] ,\quad
\omega_0 < V, 
\label{evanescent}
\end{equation}
where $k_0=[2m(\omega_0-V)/\hbar]^{1/2}$, $q_0=ik_0$, 
and $V=(V_0/\hbar)$. In the above equations the $M-$ functions are
defined as,
\begin{equation}
M(x,q,t)\equiv M(y_q)=\frac{1}{2}{\rm e}^{imx^2/2\hbar t}w(iy_q),
\label{Mosh2}
\end{equation}
where $w$ is the complex error function \cite{Abrwtz,Fdyva} $w(z)=\exp
(-z^2)erfc(-iz)$. The argument $y_q$ in Eq. (\ref{Mosh2}) 
is given by, 
\begin{equation}
y_q={\rm e}^{-i\pi /4}\left( \frac{m}{2\hbar t}\right)^{1/2}
\left[ x-\frac{\hbar q}{m}t \right],  
\label{argument}
\end{equation}
where $q=\pm k_0, \pm iq_0$.

From the analysis given in Subsection A.1, one can 
obtain the long time behavior of Eqs. (\ref{propagation})
and (\ref{evanescent}). For the long time regime $t\rightarrow \infty $
the asymptotic solutions are,
\begin{equation}
\psi_<^a(x,t)={\rm e}^{-i\omega_0 t}{\rm e}^{-q_0 x}, \quad \omega_0<V,
\label{asinto}
\end{equation}
and
\begin{equation}
\psi_>^a(x,t)= {\rm e}^{-i\omega_0 t}{\rm e}^{-ik_0 x}, \quad \omega_0>V.
\label{asintok}
\end{equation}

In what follows we shall explore the main features of waves along the 
potential region, $x>0$, generated by a source with the initial frequencies
$\omega_0 > V$ and $\omega_0 < V$.
Our emphasis will be in the second case, where we shall reveal novel aspects 
in the spatial evolution of quantum waves.

\subsection{Source with an initial frequency $\omega_0 > V$}

\begin{figure}[!tbp]
{\includegraphics[width=3.3in]{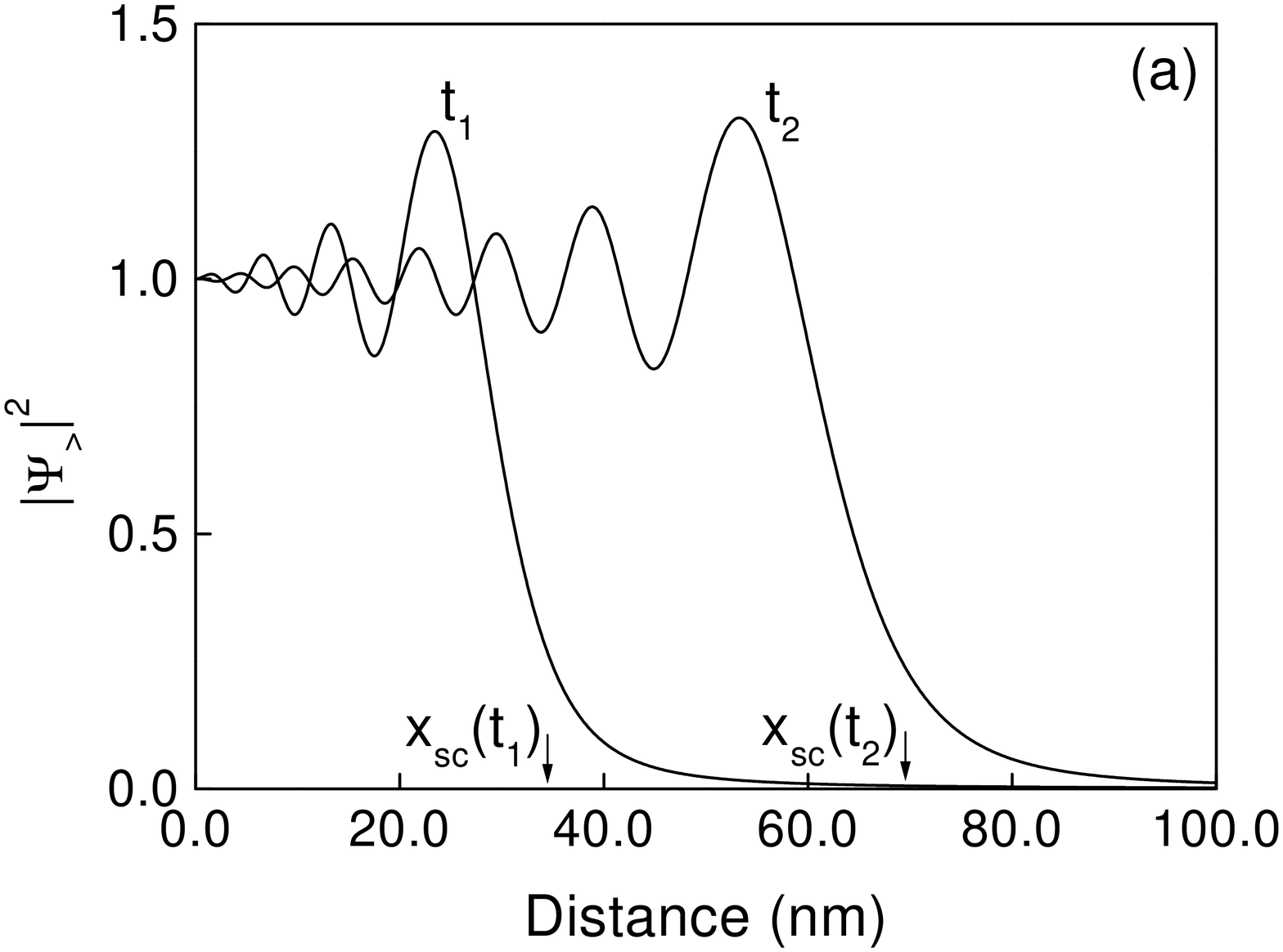}}
{\includegraphics[width=3.3in]{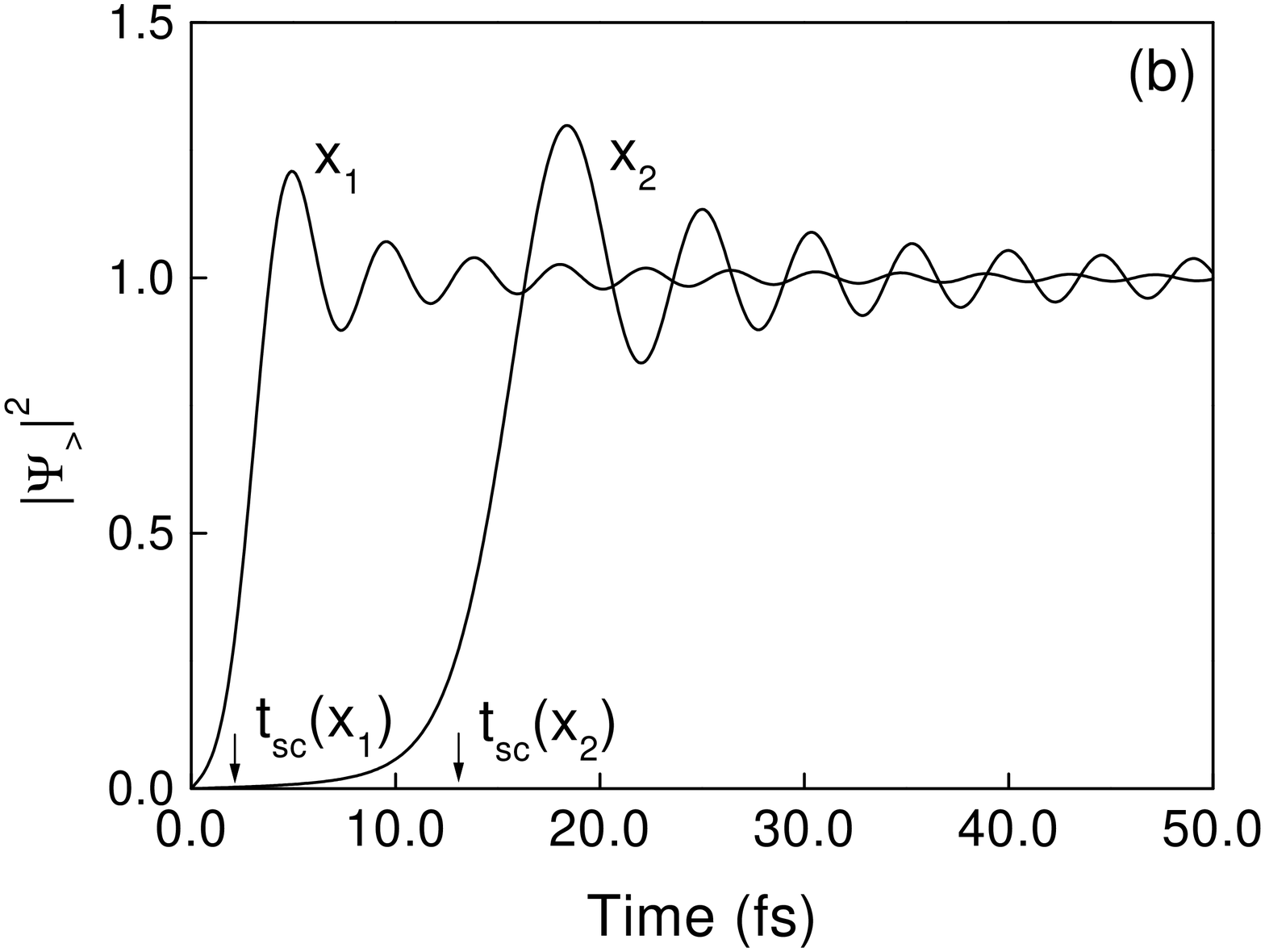}}
\caption{(a) Spatial evolution of $|\psi_{>}|^{2}$, for two different 
values of time: $t_1=15.0$ fs, and $t_2=30.0$ fs. The positions
$x_{sc}(t_i)=v_{k_0}t_{i}$ of the semiclassical fronts are indicated by
an arrow in each curve. (b) Time evolution of $|\psi_>|^2$ at different
fixed positions $x_1=5.0$ mn and $x_2=30.0$ nm, in order to show 
the wavefront propagation. The positions 
$t_{sc}(x_i)=(x_i/v_{k_0})$ of the semiclassical fronts are also indicated by
an arrow. Notice how as time increases the solution 
tends to the correct asymptotic value given by $|\psi_>^a|^2=1$.}
\label{fig1}
\end{figure}

Although the main purpose of the present work is to explore the dynamics
of the solution for a source with an initial frequency $\omega_0<V$, 
for completeness we begin our discussion by illustrating some cases for 
$\omega_{0}>V$. 
In this subsection we investigate the evolution of the solution given by Eq.
(\ref{propagation}). We choose the following parameters for the system:
height of the potential step, $V_0=1.0$ eV, and the energy of the initial
wave, $E_0=2.0$ eV. The calculated values of $|\psi_>|^2$ are shown in 
Fig. \ref{fig1} (a) as a function of the position $x$, for two fixed times
$t_1=15.0$ fs, and $t_2=30.0$ fs.
It is clearly appreciated in Fig. \ref{fig1} (a), that a main wavefront 
with damped oscillations is propagating from left to right. One can easily
verify by numerical inspection that the main wavefront travels approximately
with the semiclassical group velocity $v_{k_0}\simeq \hbar k_0/m$. The 
arrows in Fig. \ref{fig1} (a) indicate the semiclassical positions
$x_{sc}(t_i)=v_{k_0}t_i$, ($i=1,2$).

The plots of $|\psi_>|^2$ vs $t$, for a two different fixed positions
$x_1=5.0$ nm and $x_2=30.0$ exhibited in Fig. \ref{fig1} (b),
complement our discussion for this case. Here we also illustrate the 
propagation of the main wavefront for the specific positions $x_i$ 
indicated in the figure. Notice how the probability density rises 
from zero and grows monotonically towards a maximum value, from 
which it starts to oscillate until it reaches the stationary value. 
This sudden rise occurs approximately at times $t_{sc}(x_i)=(x_i/v_{k_0})$,
which correspond to the arrival of the main wavefront at the fixed position
$x_i$. It is interesting to mention that these propagation features are similar
to the {\it diffraction in time} patterns observed in free particle
propagation within the Moshinsky shutter model\cite{Moshy}, which is
in fact a different type of initial condition.
 
Although we see that in the case $\omega_0>V$ it is permissible to 
speak of a propagation of a main wave front, the issue has become 
controversial in case of initial waves with $\omega_0<V$ 
\cite{Stevens, Moretti,JPA,ButtThom,MugBu}. The results for this more
complicated case is discussed in the next subsection.

\subsection{Source with an initial frequency $\omega_0 < V$}

\begin{figure}[!tbp]
{\includegraphics[width=3.3in,height=2.0in]{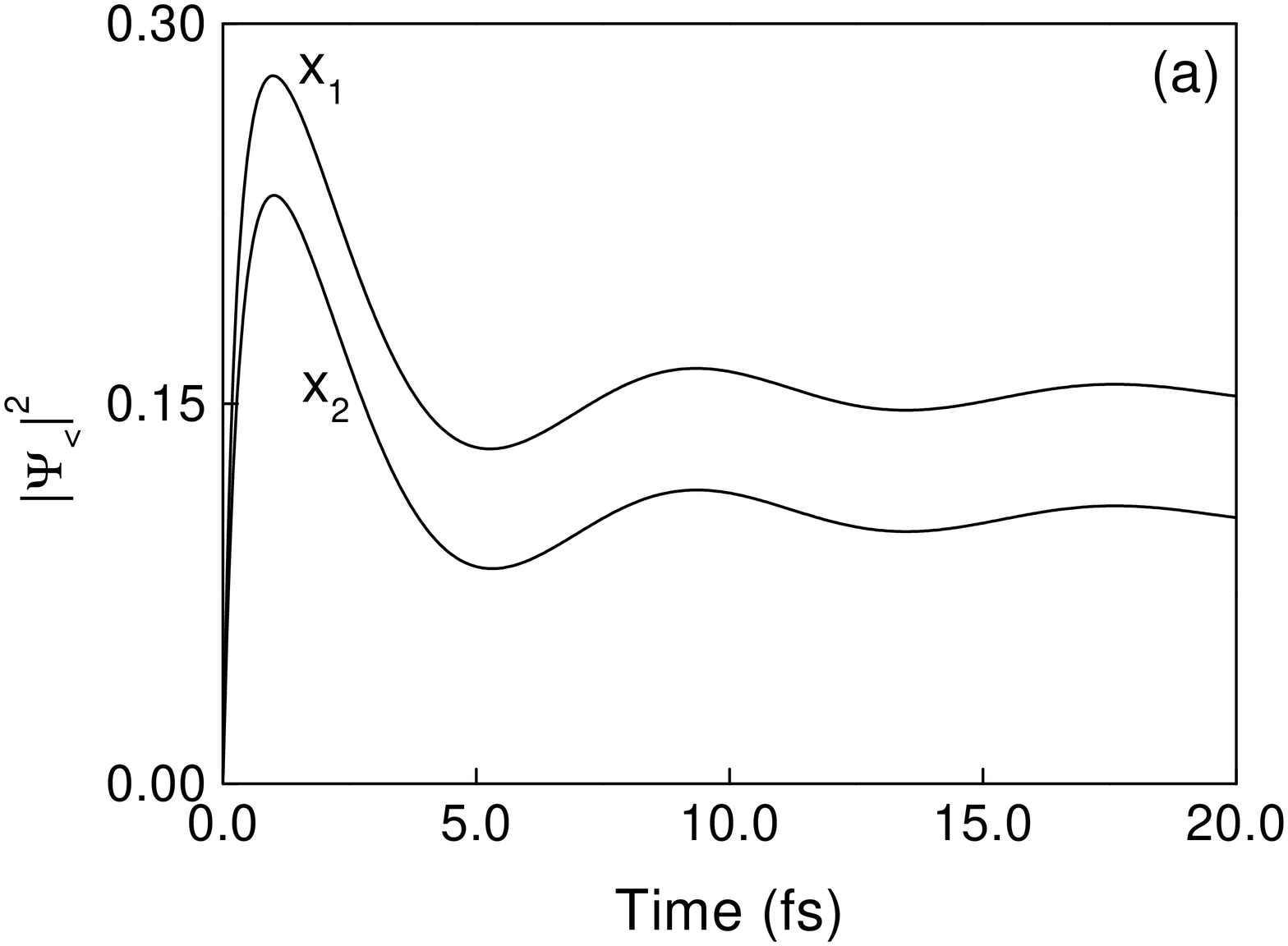}}
{\includegraphics[width=3.3in,height=2.0in]{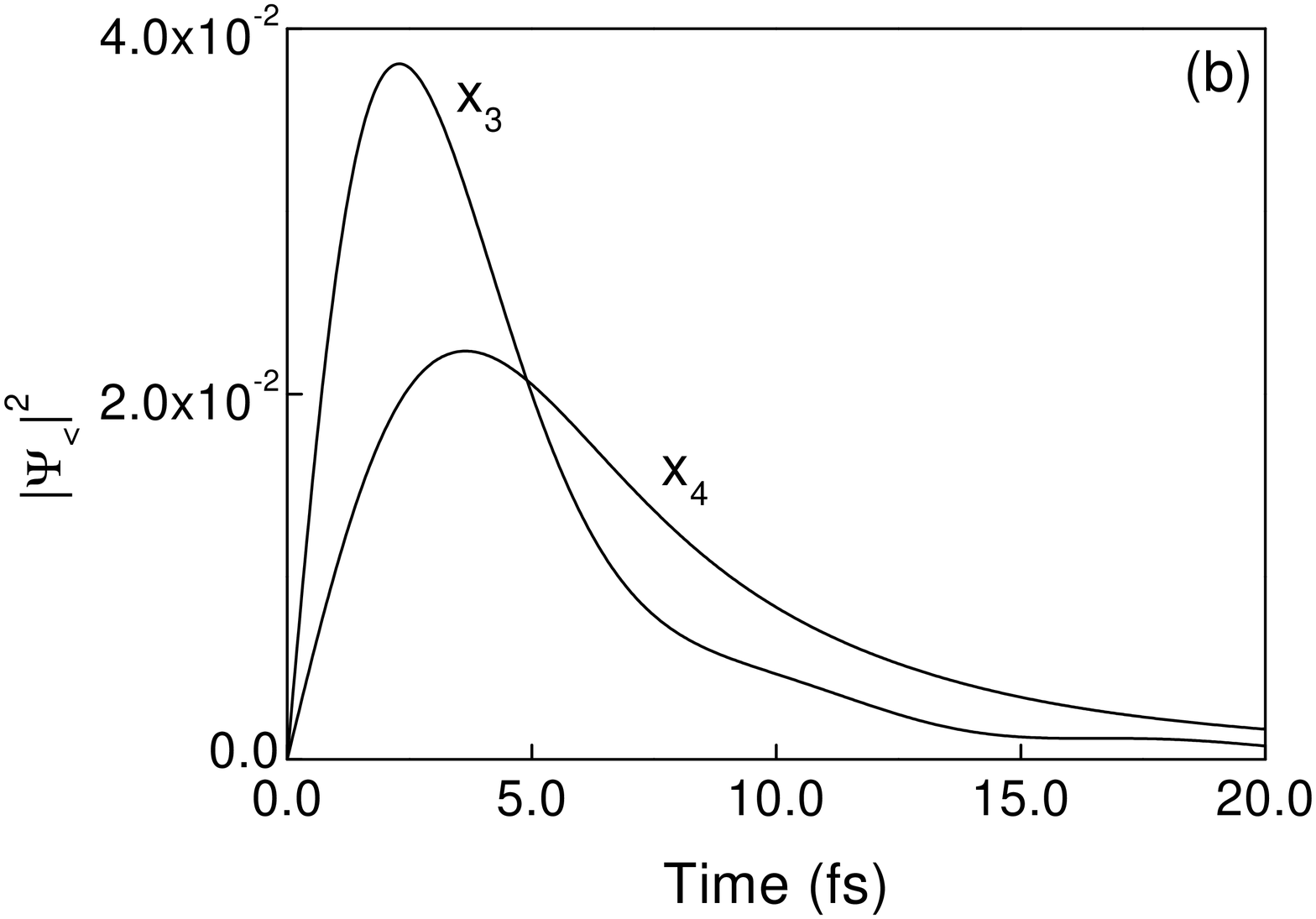}}
\caption{Time evolution of $|\psi_<|^2$, at different values
of the position along the internal region. Notice that (a) at 
the fixed positions $x_1=1.2$ nm, and $x_2=1.5$ nm, the 
probability density exhibits an oscillating behavior similar to
the diffraction in time phenomenon. At the positions (b) 
$x_3=6.0$ nm, and $x_4=10.0$ nm, located deep inside the
internal region, the oscillating pattern no longer exists, 
and it is replaced by a pulse-like structure.}
\label{fig2}
\end{figure}

In what follows we shall explore the main features of the time and
spatial behavior of the probability density 
for a source with an initial frequency $\omega_0 < V$.
We consider here the same potential parameters as in the previous 
subsection, but the energy is now chosen below the height of the 
barrier, $E_0=0.5$ eV. In Fig. \ref{fig2} (a) we show the time 
evolution of $|\psi_<|^2$ at the positions $x_1=1.2$ nm and 
$x_2=1.5$ nm. Notice that the curves exhibit a behavior similar 
to the diffraction in time pattern observed in the previous subsection.
As we shall discuss below, this behavior is not related to the 
propagation of a main wave front, and requires a different interpretation. 
In fact, the time-diffraction-like pattern disappears after a certain depth 
inside the potential, where the probability density exhibits now a 
pulse-like structure, as can be appreciated in Fig. \ref{fig2} (b). 

\begin{figure}[!tbp]
{\includegraphics[width=3.3in,height=2.0in]{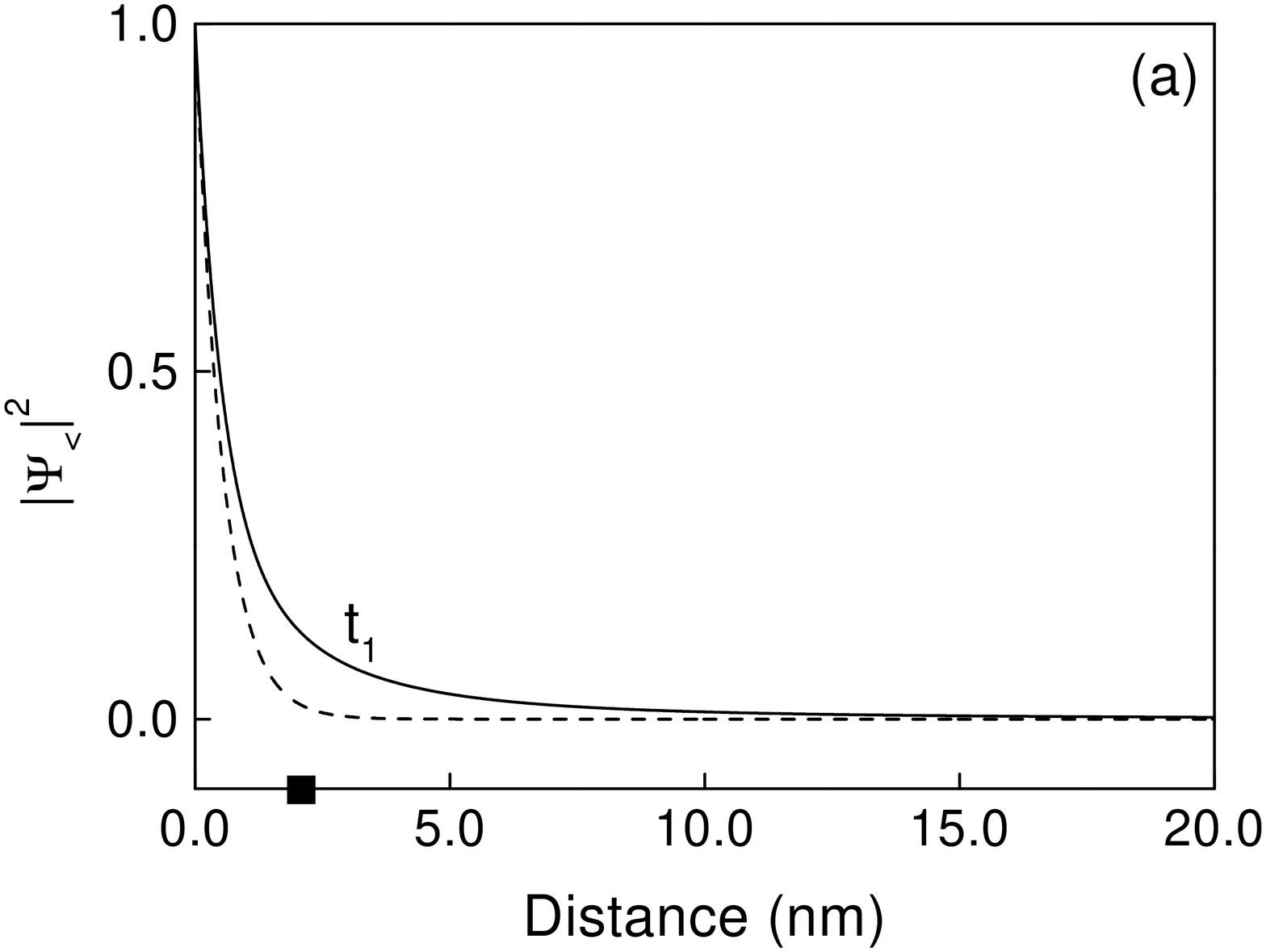}}
{\includegraphics[width=3.3in,height=2.0in]{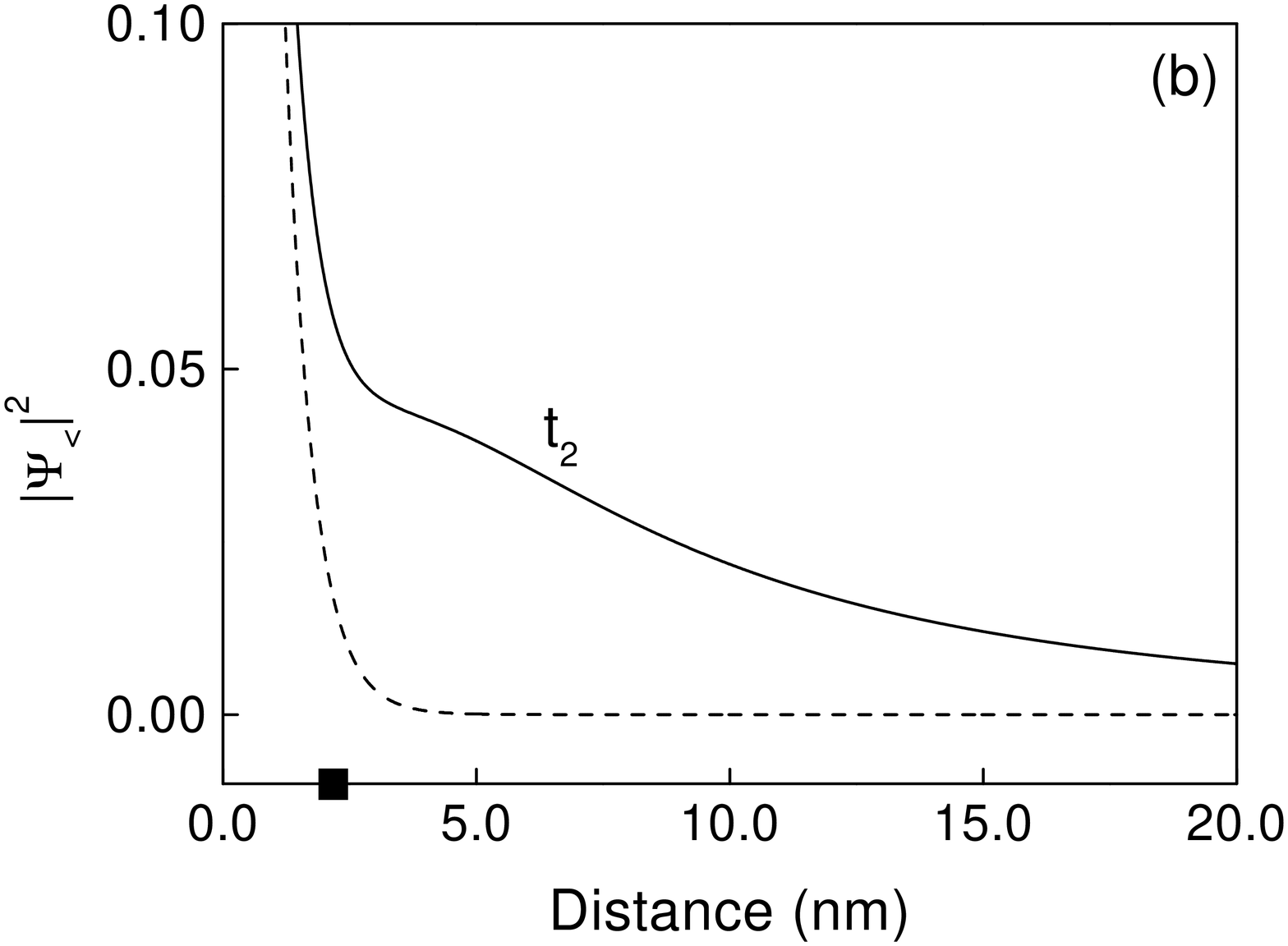}}
{\includegraphics[width=3.3in,height=2.0in]{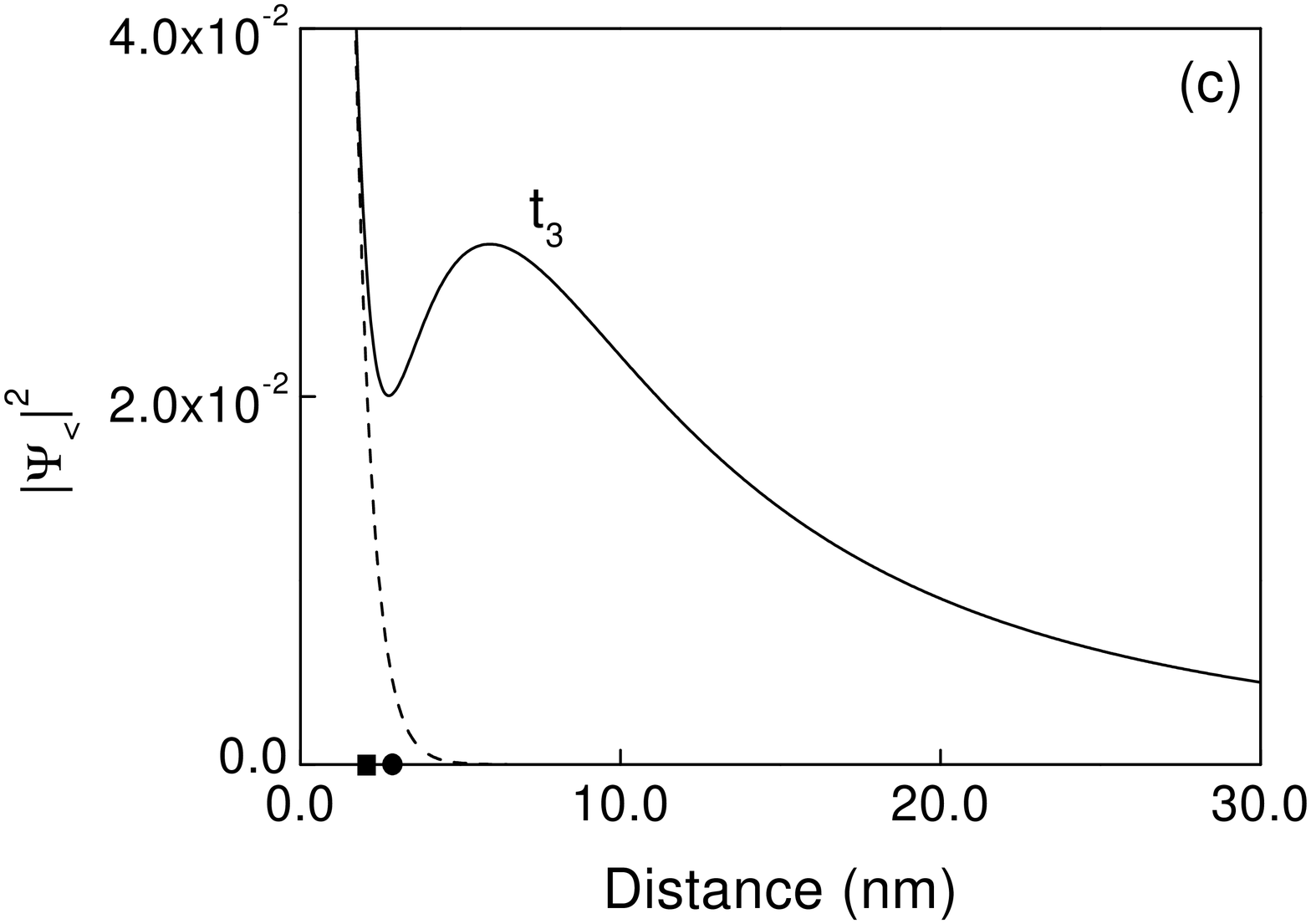}}
{\includegraphics[width=3.3in,height=2.0in]{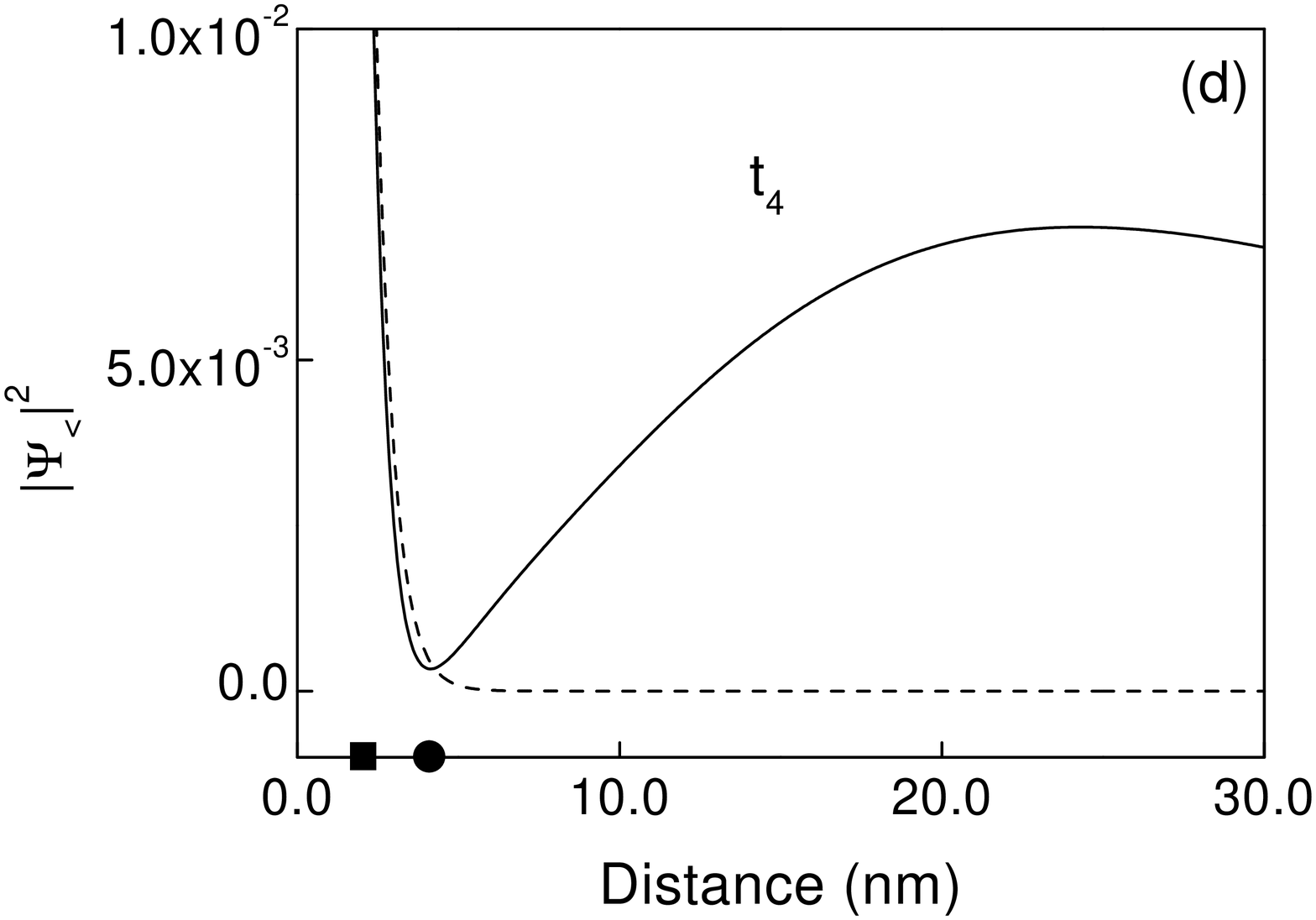}}
\caption{ Birth of the probability density $|\psi_<|^2$
(solid line) as function of the distance $x$, for increasing
values of time: $t_{1}=1.0$ fs, $t_{2}=3.0$ fs, $t_{3}=4.0$ fs,
and $t_{4}=15.0$ fs. In part (a), we see a monotonically 
decreasing curve. However, a bump begins to appear in (b),
which corresponds to the birth of a {\it transient pulse}. In (c) and (d),
the maximum of this pulse propagates along $x>X_0$.  
In all cases, the stationary solution $|\psi_<^a|^2$ (dashed line)
is included for comparison. A full square and a dot indicate the position
of $X_0=2.134$ and $X_R$, respectively.}
\label{fig3}
\end{figure}

In order to understand the transition between the quite different 
dynamical behavior of $|\psi_<|^2$ observed in Fig. 
\ref{fig2} (a) and (b), a series of snapshots of $|\psi_<|^2$
vs $x$ are shown in Fig. \ref{fig3} (solid lines) for increasing
values of time. We can appreciate in Fig. \ref{fig3} (a) that at 
short times the probability density behaves like the stationary 
solution $|\psi_<^a|^2$ (dashed line) along the internal 
region. However, the probability density, $|\psi_<|^2$, 
dramatically changes as time elapses. This is depicted in  
Fig. \ref{fig3} (b), where we see the appearance of a bump 
for values of the position $x$ larger than $X_0$. This special
point $X_0$ corresponds to the onset of a pulse-like structure
that we call {\it transient pulse} (forerunner). In Figs. \ref{fig3} (c) 
and (d), the transient pulse is now fully formed, and propagates
along the region $x>X_0$. However, notice that as time increases, 
the birth of the pulse occurs at values of the position $X_R$, 
greater than value of the lower-bound, $X_0$.
The positions $X_0$ (full square) and $X_R$ (full dot) are indicated
in Fig. \ref{fig3}. 
We have systematically studied this transition for different potential
steps, as well as for different energies of the initial waves, and
our results indicate that the onset of the {\it transient pulse} always occurs
at a specific depth, $X_0$, inside the potential, which is twice the value of 
the characteristic {\it penetration length} $x_p=q_0^{-1}$ of the stationary 
solution: $X_0\approx 2x_p$. 
The existence of this two regimes for the point source model has also been 
reported in Ref. \cite{gvdm02}. A similar separation into two regimes, but 
in different models has been observed by one of the authors in Refs.
\cite{JPA,gcv01}.

We shall later see that the approximate value of $X_R$ 
can be determined approximately by using a simple analytical formula. 
In what follows we shall explore in more detail the regimes $0<x<X_0$,
and $x>X_0$. We shall refer to the latter as the {\it transient pulse regime}.

\subsubsection{ $0<x<X_0$ regime}

\begin{figure}[!tbp]
{\includegraphics[width=3.3in]{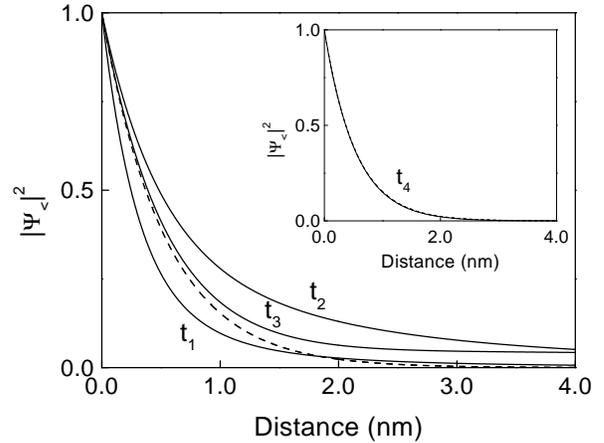}}
\caption{Plot of $|\psi_{<}|^{2}$ as a function of the position at
different values of time: $t_{0}=1.0$ fs (solid line) $t_{1}=2.0$ fs
(dotted line), and $t_{2}=3.0$ fs (dashed-dotted line).
The stationary solution $|\psi_{<}^{a}|^{2}$ (dashed line)
is also included for comparison. The inset shows $|\psi_{<}|^{2}$ 
(solid line) at a later time $t_{3}=15.0$ fs, which becomes indistinguishable
from the stationary solution (dashed line). }
\label{fig4}
\end{figure}

This regime corresponds to short distances inside the potential, 
namely, to positions $x$ such that $0<x<X_0$. Here we shall explain the
oscillatory  patterns of the $|\psi_<|^2$ vs $t$ plots, observed within 
this finite interval and exhibited in Fig. \ref{fig2} (a). We shall see that, 
rather than originated by the passage of a traveling wavefront 
(as it occurs in the case $\omega_0 > V$), this effect is due to 
fluctuations of the probability density.
This 
dynamical behavior is illustrated in Fig. \ref{fig4}, where
we present a series of snapshots of $|\psi_<|^2$ vs $x$, 
for increasing values of time indicated in the figure.
Here we have included for comparison
the calculation of the stationary probability density
$|\psi^a_{<}|^{2}$ (dashed line). Notice that the 
evolution of the time dependent probability density occurs 
in such a way that $|\psi_<|^2$ is found
sometimes above or below the asymptotic value. This illustrates
that, in this regime, the probability density fluctuates around
the stationary solution before reaching its asymptotic limit.
At long enough times, we must have
$|\psi_{<}|^{2}\rightarrow|\psi^a_<|^2$, which is illustrated
in the inset of Fig. \ref{fig4}, where for $t=15.0$ fs both 
curves become indistinguishable.
\begin{figure}[!tbp]
{\includegraphics[width=3.3in]{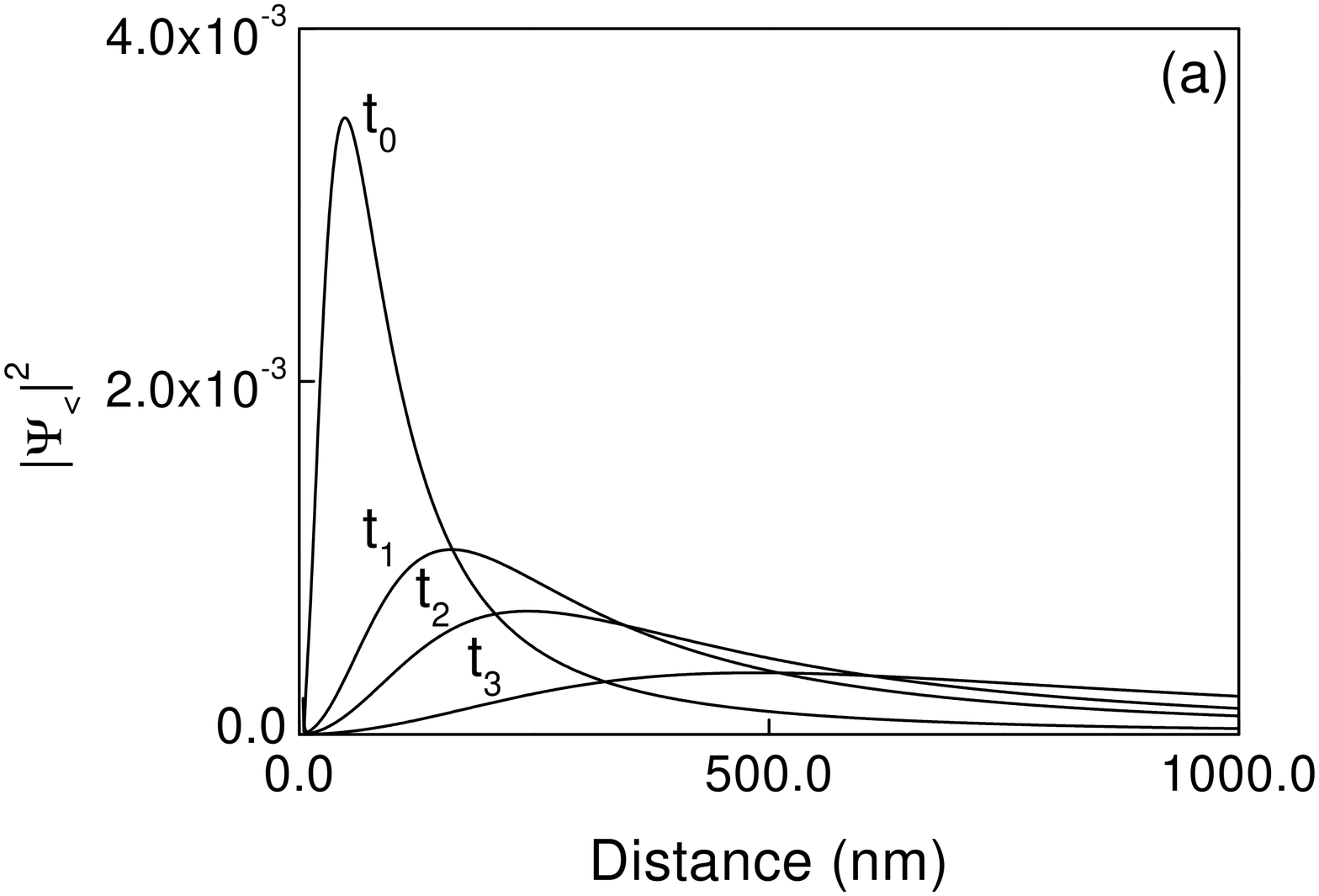}}
{\includegraphics[width=3.3in]{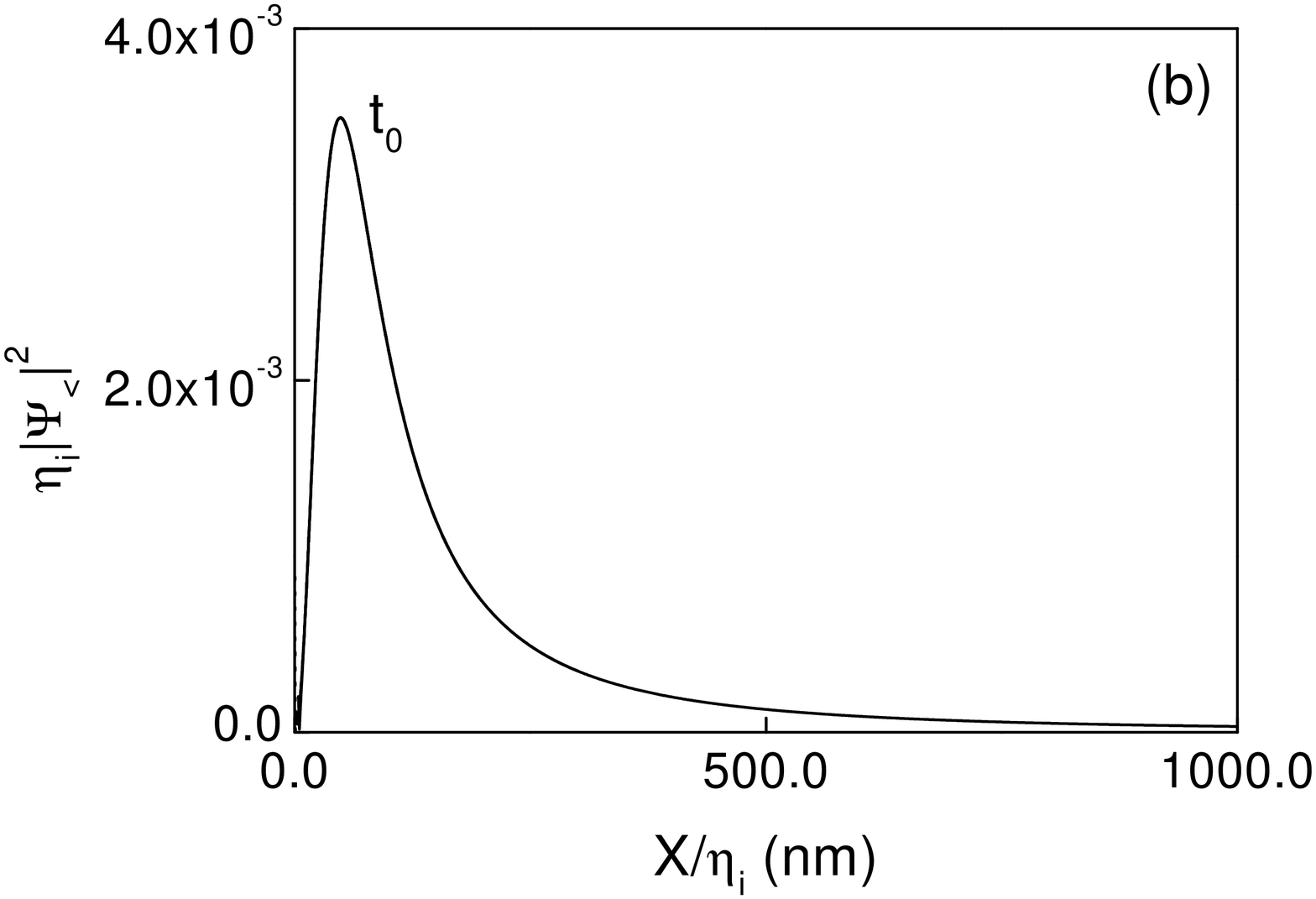}}
\caption{(a) Spatial evolution of the {\it transient pulse} calculated from
$|\psi_<|^{2}$ for increasing values of time: $t_{0}=30.0$ fs,
$t_{1}=100.0$ fs, $t_{2}=150.0$ fs, and $t_{3}=300.0$ fs. Notice
the different values of the pulse peak and width for each particular
time. (b) Plot of the rescaled probability density
$\eta_{i}|\psi_{<}|^{2}$ ($i=1,$ $2,$ $3$) as function of a
variable $x/\eta_i$, for the same cases 
depicted in (a). All curves become indistinguishable and coincide 
with the $|\psi_{<}|^{2}$ vs $x$ plot of the case $t=t_0$. Also
included here is a plot of Eq. (\ref{invariant}), which, as expected,
coincides exactly with all the curves.}
\label{fig5}
\end{figure}

The above results are relevant since they clearly show that 
we can not associate in this regime a semiclassical group velocity 
to the main wave-front towards its steady state. In other words 
it is not possible to define a semiclassical main front attenuated 
exponentially by $e^{-q_0 x}$, traveling along the internal region 
with a group velocity $v_{q_0}=(\hbar q_0/m)$, as suggested by Moretti 
\cite{Moretti}.  However, a special kind of propagation actually exists
for $x>X_0$ (opaque barrier conditions), which consists of the 
traveling forerunner mentioned in the Section I. It should be 
emphasized however that this transient structure (in view of its 
frequency content) is not associated to a tunneling process\cite{MugBu}.
We shall show in the next subsection that the maximum of this {\it forerunner} 
travels at exactly the group velocity, $v_{q_0}$. 

\subsubsection{Transient pulse regime ($x>X_0$)}

This regime corresponds to the case $x>X_{0}$, where the dynamics of the 
probability density $|\psi_<|^2$ is governed by a {\it transient pulse}.
This transient structure originates from  $X_{0}$ onwards, and evolves 
thereafter along $x>X_{0}$, as depicted in figures \ref{fig3} (c) and
\ref{fig3} (d). In Fig. \ref{fig5} (a), we plot $|\psi_<|^2$ vs $x$ for 
the increasing values of time: $t_{0}=30.0$ fs, $t_{1}=100.0$ fs,
$t_{2}=150.0$ fs, and $t_{3}=300.0$ fs. Here, no traveling main front
is observed, as in the $\omega_0 > V$ case. Instead, we see a forerunner that 
broadens as it travels along the internal region, with diminishing 
amplitude, similar to the propagation and spreading of a wavepacket. 

In addition to the above, the evolution of the probability density
in this regime, has an interesting scaling property. If we
express the times $t_{i}$ ($i=1,$ $2,$ $3$) as multiples of $t_{0}$, namely
$t_{i}=\eta_{i}t_{0}$ (in our examples: $\eta_{1}=3.333$, $\eta _{2}=5.0$,
and $\eta_3=10.0$), and use the factors $\eta_{i}$ to renormalize
both axis, in such a way that we now plot $\eta_{i}|\psi_<|^2$ vs $(x/\eta_i)$,
a striking result is obtained: all curves coincide. This invariance under
scaling is illustrated in Fig. \ref{fig5} (b), where all the curves
corresponding to $i=1,2,3$, are indistinguishable among them, coinciding
with the curve $|\psi_<|^2$ vs $x$ corresponding to $t_{0}$. 

The scaling property illustrated above by means of a numerical example,
can also be explained using our analytic formulas. However, instead of
using the formal solution given by Eq. (\ref{evanescent}), we shall 
first derive a simple expression for the spatial evolution of $|\psi_<|^2$.
Notice that the behavior of $|\psi_<|^2$ is essentially
governed by two $M$ functions, namely, $M(y_{iq_0})$ and  $M(y_{-iq_0})$. 
By a numerical inspection we find that the series of the $M$ functions 
discussed in Subsection 1 of Appendix A, apply in this regime. 
In order to exploit the analytical properties of the Moshinsky functions,
first we have to determine the location in the complex-plane of the 
phases $\phi_{\pm iq_0}$ of the arguments $y_{\pm iq_0}$. We find that 
the phase $\phi_{-iq_0}$, is always within the interval
$-\pi /2<\phi_{-iq_0}<\pi /2$. Thus the series representation given by 
Eq. (\ref{m1}) applies in this case. 
Similarly, we find that although the phase $\phi_{iq_0}$, may change from
$\pi /2<\phi_{iq_0}<3\pi /2$ to $-\pi /2<\phi_{iq_0}<\pi /2$ in the 
vicinity of the maximum of the {\it transient pulse}, the series representation 
of the $M$ functions given by Eq. (\ref{m2}) gives an excellent 
approximation. Therefore, we find that in this regime, the
behavior of the probability density is essentially governed by the
first terms of the series representation of Eqs. (\ref{m1}) and
(\ref{m2}). That is,
\begin{equation}
M(y_{-iq_0})\simeq \frac{1}{2}e^{imx^{2}/2\hbar t}\left[ \frac{1}{\pi
^{1/2}y_{-iq_{0}}}\right], 
\label{masint1}
\end{equation}
and
\begin{equation}
M(y_{iq_{0}})\simeq \frac{1}{2}e^{imx^{2}/2\hbar t}\left[
2e^{y_{iq_{0}}^{2}}+\frac{1}{\pi ^{1/2}y_{iq_{0}}}\right]. 
\label{masint2}
\end{equation}
After substituting the above expressions in Eq. (\ref{evanescent}),
and performing a few algebraical manipulations, we obtain the
following expression for the probability density,
\begin{equation}
\psi_<\simeq \psi _{<}^{a}+\psi_{tp},
\label{evanasin}
\end{equation}
where $\psi_{<}^{a}$ is the stationary solution given by Eq. 
(\ref{asinto}), which decays exponentially along the internal
region. Here $\psi_{tp}$ stands for,
\begin{equation}
\psi_{tp}=\frac{1}{2\pi ^{1/2}}e^{i(mx^{2}/2\hbar t-Vt)}\left[ \frac{1}{%
y_{iq_{0}}}+\frac{1}{y_{-iq_{0}}}\right],
\label{tranpulse}
\end{equation}
which as we shall see below, accurately describes the {\it transient pulse}.
We have found by numerical inspection that Eq. (\ref{evanasin}) gives an excellent
description of the spatial evolution of $|\psi_<|^2$, provided that
$t>(X_0/v_{q_0})$. The result given by Eq. (\ref{evanasin}) shows us 
that the probability density can be described by a subtle interplay between
$\psi_{<}^{a}$ and $\psi_{tp}$. 
A measure of such an interplay can be given by the ratio $R=|\psi_<^a/\psi_{tp}|^2$.
We have found that the observed transition discussed in Fig. {\ref{fig3}, 
occurring at $X_R$, corresponds to values of the position such that condition
$R=1$ is satisfied. 
 
If we now focus our attention in describing the dynamics of the pulse, from Eq.
(\ref{tranpulse}) we can obtain a simple analytical expression for this 
transient structure, namely,
\begin{equation}
|\psi_{tp}(x,t)|^{2}=\frac{2}{\pi }\frac{(\hbar x^{2}t/m)}{%
\left[ x^{2}+(\hbar q_{0}t/m)^{2}\right] ^{2}}.
\label{tfe1}
\end{equation}
The analysis from Eq. (\ref{masint1}) to Eq. (\ref{tfe1}),  
is equivalent to the one performed in Ref. \cite{MugBu}
in terms of a pole and a saddle-point contribution, with a 
different notation and terminology.
It can be appreciated in Fig. \ref {fig6} (a), that Eq. (\ref{tfe1})
provides an excellent description for the propagating pulse. Here,
we have plotted $|\psi_{tp}| ^{2}$ (solid line) as a function of time
at a fixed position $x=8.0$ nm, and the curve almost coincide with the
calculated with the formal solution, Eq. (\ref{evanescent}) (dashed line).
We also compare Eqs. (\ref{evanescent}) and (\ref{tfe1}) with a plot
of the probability density as a function of position, see Fig. \ref {fig6}(b),
and an the description of our approximate formula is excellent, since both
calculations are indistinguishable in the graph. 
\begin{figure}[!tbp]
{\includegraphics[width=3.3in]{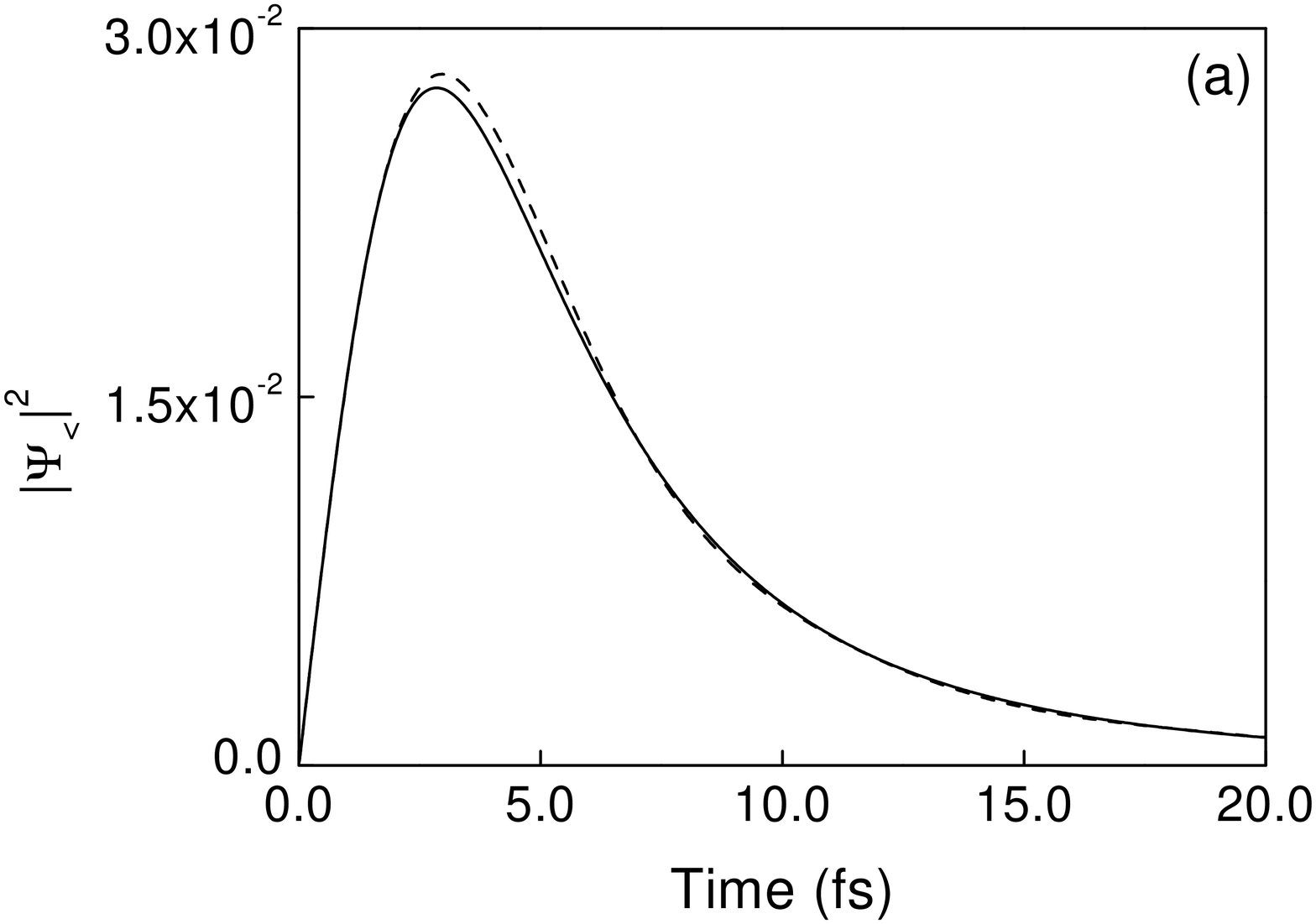}}
{\includegraphics[width=3.3in]{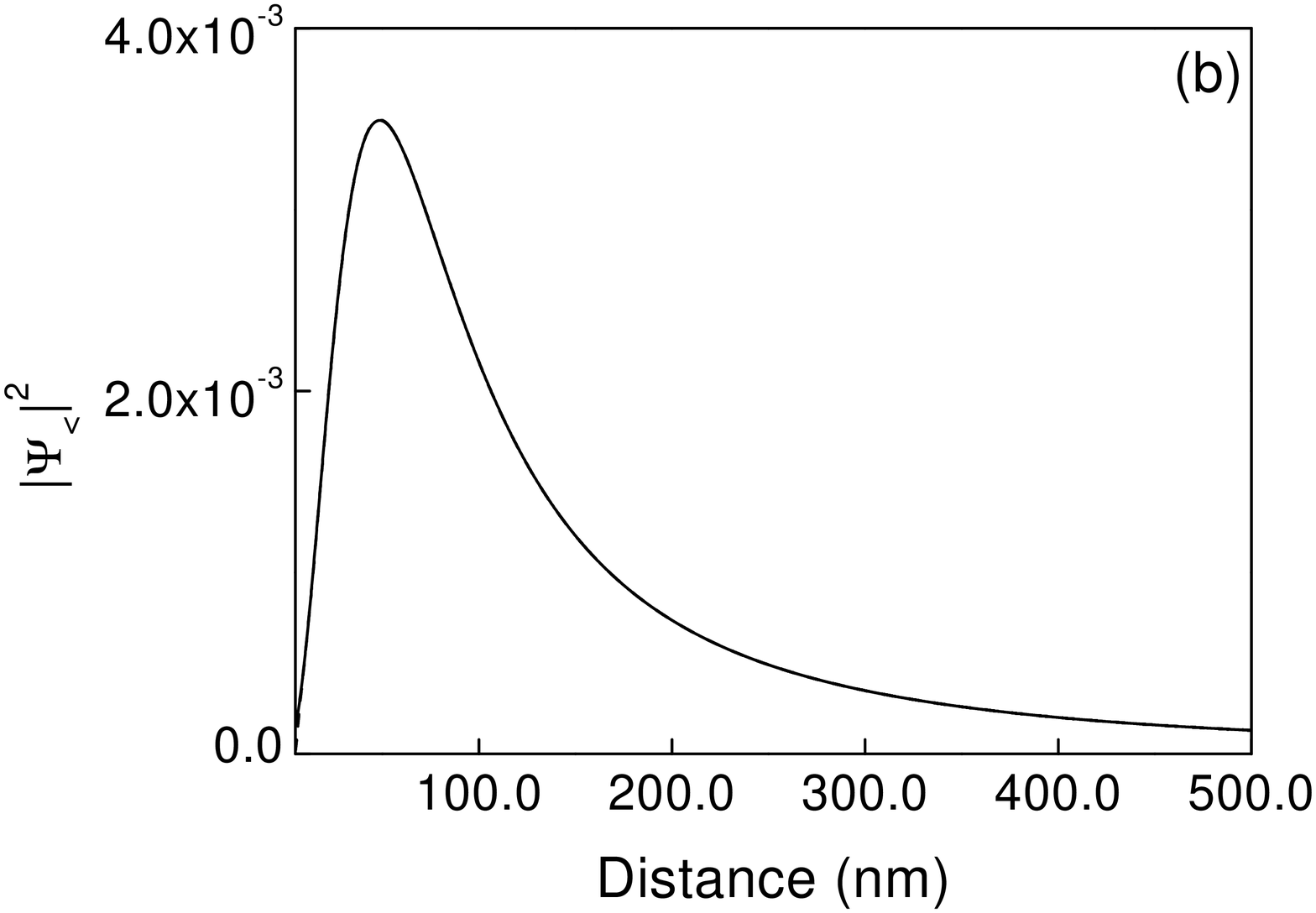}}
\caption{Comparison of $|\psi_{tp}|^{2}$ (solid line) and $|\psi_<|^2$
(dashed line), (a) as a function of time, for $x_0=8.0$ nm, and (b) as
a function of position, for $t_{0}=30.0$ fs; in this case,
both curves overlap, and become indistinguishable among them.}
\label{fig6}.
\end{figure}

In order to show the scaling property, we make in Eq. (\ref{tfe1})
the general substitution of $t^{\prime}=\eta t$ and $x^{\prime}=\eta x$.
The result of this rescaling is the following equation, 
\begin{equation}
|\psi_{tp}(x^{\prime},t^{\prime})|^{2}=\frac{2}{\eta \pi }%
\frac{(\hbar x^{2}t/m)}{\left[ x^{2}+(\hbar q_0t/m)^{2}\right]^{2}}.
\label{invariant}
\end{equation}

A plot of Eq. (\ref{invariant})
is also included in Fig. \ref{fig5} (b), which as expected becomes also 
indistinguishable from the rest of the curves. From the comparison of 
Eq. (\ref{tfe1}) and Eq. (\ref{invariant}), the scaling property can 
be written as,
\begin{equation}
\eta|\psi_{tp}(\eta x,\eta t)|^{2}=|\psi_{tp}(x,t)|^{2}.
\label{scaling}
\end{equation}
 
In what follows we shall apply Eqs. (\ref{tfe1}) and (\ref{scaling})
to analyze the speed of the propagating pulse. In particular, we contrast
the propagation of the peak value of the probability density obtained by
deriving it with respect to time, to the peak value obtained by deriving
with respect to position. As we shall show below, the corresponding speeds 
differ by a factor of $3^{1/2}$. 

The maximum value of $|\psi_{tp}|^2$ at any given point 
can be obtained straightforwardly by deriving Eq. (\ref{tfe1}). 
However, the following results show that the calculation of 
a maximum of $|\psi_{tp}|^2$ critically depends on whether 
the spatial or time evolution of the probability density is 
analyzed. From $(d/dt)|\psi_{tp}|^2=0$, we obtain the critical
time which maximizes $|\psi_{tp}|^2$ at $x_f$, the result is,
\begin{equation}
t_m=\frac {\tau}{\sqrt3},
\label{tm}
\end{equation}
where $\tau=(x_f/v_{q_0})$. This result is in exact agreement
with the scale found recently by Muga and B\"{u}ttiker \cite{MugBu} 
for the time of arrival.

It is clear that the pair $(x_f,t_m)$ gives the maximum value in a 
$|\psi_{tp}|^2$ vs $t$ plot. However, it is easy to show that the 
value $x_f$ does not correspond to the position of the peak of the 
traveling pulse. The latter, which is the maximum of a 
$|\psi_{tp}|^2$ vs $x$ plot, is in fact located at the left of 
$x_f$ (see below). From $(d/dx)|\psi_{tp}|^2=0$, at a fixed 
time $t=t_f$, we obtain the maximum of the {\it transient pulse}
as it propagates along the $x$ coordinate, it becomes,
\begin{equation}
x_m=v_{q_0}t_f.
\label{xm}
\end{equation}
Notice that for the particular case $t_f=t_m$, the maximum of the 
pulse $x_m$ is retarded with respect to $x_f$ {\it i.e.} 
$x_m=3^{-1/2}x_f$. In other words, the maximum intensity of 
$|\psi_{tp}|^2$ is attained at $x_f$, before the arrival of the 
pulse's peak takes place. Although the above statement seems to 
be in contradiction, we shall show below that these two types of
maximum values have different meanings, and consequently there 
is no contradiction at all. The values of the probability density 
at these two positions are illustrated on the $|\psi_{tp}|^2$ vs
$x$ plot (solid line) depicted in Fig. \ref{velocities}; they 
are compared at the same time, $t_m=30$ fs. The values of 
$|\psi_{tp}|^2$ at $x_m$ and $x_f$ are indicated on the curve by
a full dot and a hollow circle, respectively. Equation (\ref{xm}) 
shows that the maximum of the pulse (full dot) travels at exactly 
the semiclassical group velocity $v_{q_0}$, while the other 
(hollow circle), according to Eq. (\ref{tm}), travels faster than
the former, at the speed $3^{1/2}v_{q_0}$. In fact, the observed 
spreading of the pulse is due to the fact that all the points of
the $|\psi_{tp}|^2$ vs $x$ curve are traveling with different 
velocities.

It is clear from Fig. \ref{velocities} that the height of the 
hollow circle $h_{hc}(t_m)$ is smaller than the corresponding 
height of the full dot, $h_{fd}(t_m)$. However, the height of 
the pulse diminishes at such a rate, that when the full dot
reaches $x_f$ at a later time $t^{\prime}=3^{1/2}t_m$, its 
height $h_{fd}(t^{\prime})$ is smaller than $h_{hc}(t_m)$. 
This is illustrated in Fig. \ref{velocities} where we plot 
$|\psi_{tp}|^2$ as a function of the position $x$, at a time
$t=t^{\prime}$ (dashed line). The above effect can be easily 
shown by using the scaling property, Eq. (\ref{scaling}). 
But first, lets us obtain the values of the heights $h_{hc}$ 
and $h_{fd}$ measured at $t=t_m$, namely,
\begin{equation}
h_{hc}(t_m)=|\psi_{tp}(x_f,t_m)|^2=\frac{3}{4} \left ( \frac{1}{2\pi }%
\frac{1}{x_m q_0}\right ); 
\label{ht}
\end{equation}
and
\begin{equation}
h_{fd}(t_m)=|\psi_{tp}(x_m,t_m)|^2=\left ( \frac{1}{2\pi }%
\frac{1}{x_m q_0}\right ). 
\label{hm0}
\end{equation}
If we now feed $\eta=3^{1/2}$ in Eq. (\ref{scaling}), using the
fact that $t^{\prime}=3^{1/2}t_m$ and $x_f=3^{1/2}x_m$, we obtain
that $3^{1/2}h_{fd}(t^{\prime})=h_{fd}(t_m)$. This
allows us to write the height ratio as,
\begin{equation}
\frac {h_{hc}(t_m)}{h_{fd}(t^{\prime})}=\frac {3\sqrt3}{4}>1.
\label{ratio}
\end{equation}
Thus, we have in general demonstrated that the maximum value of 
$|\psi_{tp}|^2$, measured at a fixed position $x_f$, is always 
reached there before the arrival of the maximum of the transient 
forerunner at that particular position. In view of this result, 
the role of the different time scales is now clear. While the 
traversal time $\tau$ does in fact correspond to the time of 
arrival of the maximum of the transient forerunner at $x_f$, the
time scale $3^{-1/2}\tau$ corresponds to the attainment of the maximum 
value of $|\psi_{tp}(x_f,t)|^2$ at $x_f$, which as demonstrated 
above, is always associated to values of the probability density
located at the front tail of the forerunner.

The above discussion shows that the calculation of the time of arrival
may critically depend on whether the spatial or the time evolution 
of the probability density is analyzed. Although the interpretation 
of the results is different in each case, we believe that both kind 
of analysis are important and complementary.
It is important to point out that a time-frequency analysis
of the {\it forerunner}, performed in Ref. \cite{MugBu},  
has revealed that it is composed predominantly by over-the-barrier
(non-tunneling) frequencies.

\begin{figure}[!tbp]
{\includegraphics[width=3.3in]{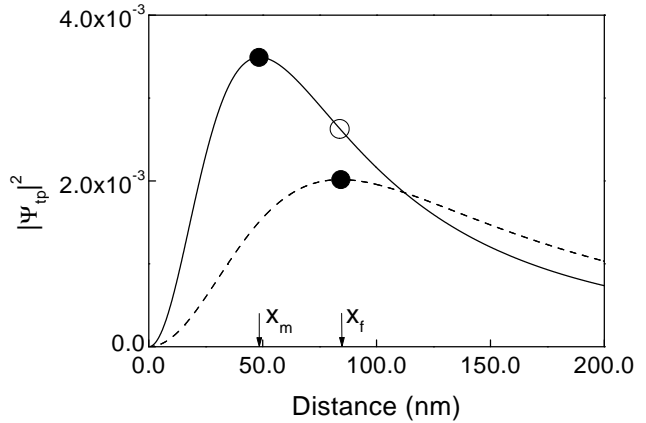}}
\caption{Spatial evolution of the {\it transient pulse} calculated from
$|\psi_{tp}|^{2}$ for $t_{m}=30.0$ fs (solid line). Its values at the positions
$x_m=48.639$ nm and $x_f=84.246$  nm are indicated respectively by a full dot 
and a hollow circle on the solid line. At a later time, $t^{\prime}=3^{1/2}t_m$, 
the maximum of the forerunner reaches the position $x_f$, as indicated by the full 
dot on the dashed line. Notice the reduction of its height}
\label{velocities}
\end{figure}

\section{Conclusions}

The issue of spatial and time dependence of quantum waves 
in a step potential barrier has been analyzed by means of an 
exact solution to the time-dependent Schr\"{o}dinger equation, 
for a point source initial condition.
The main results of our study are the following:
(1) For the case $\omega_0< V$ in the opaque
barrier regime ($xq_0\gg 1$), we found that the peak of the
{\it forerunner} travels at exactly the group velocity 
$v_{q_0}=(\hbar q_0/m)$, which implies that the 
time of arrival of the maximum of the peak 
at a fixed position in space $x_f$, is given by the traversal 
time $\tau=(x_f/v_{q_0})$. This is in contrast to the time 
scale $3^{-1/2}\tau$ described in Ref. \cite{MugBu}.
(2) We derived from the formal solution a closed analytical 
formula that accurately describes the dynamics of the 
{\it forerunner}, and found that it obeys an interesting and 
useful scaling property.
(3) Although $3^{-1/2}\tau$ actually corresponds to the maximum 
of the {\it forerunner} when it is analyzed as a function of time
(and fixed $x$), we demonstrated by means of the scaling property 
that this time scale always correspond to the passage of the front 
tail (not the maximum) of the forerunner across the fixed position
$x_f$.

 As a final remark, it is important to point out that the 
{\it forerunner} is mainly composed by over-the-barrier (non-tunneling)
frequencies, as recently shown by Muga and B\"uttiker \cite{MugBu}. 
Thus, it is a surprising fact that the traversal time $\tau$, usually 
regarded as one of the possible tunneling times, is here associated 
to a non-tunneling process.

\begin{acknowledgments}
The authors, J. V. and R. R., acknowledge financial support from
Conacyt, M\'{e}xico, through Contract No. 431100-5-32082E. The 
authors thank Alberto Hern\'{a}ndez for useful discussions.
\end{acknowledgments}

\appendix

\section{POINT SOURCE PROBLEM}

In this appendix we shall present a different derivation of the solutions
to the point source problem \cite{MugBu,Moretti}, by means of the Laplace 
Transform method. Let us consider an exact analytical solution to the 
time-dependent Schr\"{o}dinger equation for a step potential barrier 
$V(x)=\Theta (x)V_0$,
\begin{equation}
\left[ -\frac{\hbar ^{2}}{2m}\frac{\partial ^{2}}{\partial x^{2}}%
+V(x)-i\hbar \frac{\partial }{\partial t}\right] \psi (x,t)=0,  
\label{step}
\end{equation}
subject to the source boundary condition given by Eq. (\ref{ci}).
We must also assume that inside the barrier ($x\geq 0$) the initial state 
$\psi_0$\ satisfies the boundary condition $\psi_0\left( x;t=0\right)=0$, 
for $x\geq 0$.

To obtain the solution for $x>0$ at $t>0$, we begin by Laplace transforming
equation (\ref{step}) using the standard definition
\begin{equation}
\overline{\psi }(x;s)=\int\nolimits_{0}^{\infty }\psi (x,t)e^{-st}dt,
\label{lapt}
\end{equation}
with the initial condition given by Eq. (\ref{ci}). Let us first consider
the  case propagating waves ($\omega_0 >V$) which corresponds
to a dispersion relation $k_0=[\beta (\omega_0-V)]^{1/2}$, where we have
defined $\beta=(2m/\hbar )$ and $V=(V_0/\hbar)$. The Laplace transformed 
solution reads, 
\begin{equation}
\overline{\psi }(x;p)=c_{1}e^{ipx},\qquad x\geq 0,  
\label{sol1}
\end{equation}
where $p^2=\beta (is-V)$. The corresponding Laplace transform of Eq. (\ref
{ci}) yields
\begin{equation}
\overline{\psi}_0(0;s)=\frac{1}{s+i\omega_0},\qquad t>0.  
\label{citl}
\end{equation}
By combining Eqs. (\ref{sol1}) and (\ref{citl}) evaluated at $x=0$ we can
determine the value of the constant $c_1$, which yields the Laplace
Transformed solution for $x\geq 0$,
\begin{equation}
\overline{\psi }(x;p)=\frac{e^{ipx}}{(p+k_{0})(p-k_{0})},\qquad x\geq 0.
\label{laptrans}
\end{equation}

The time dependent solution for $x>0$ is readily obtained by performing the
inverse Laplace transform of Eq. (\ref{laptrans}), using the Bromwich
integral formula,
\begin{equation}
\psi (x,t)=\frac{1}{2\pi i}\int\limits_{\gamma ^{\prime }-i\infty }^{\gamma
^{\prime }+i\infty }\overline{\psi }(x;s)e^{st}ds,  
\label{bromwich1}
\end{equation}
where the integration path is taken along a straight line $s=\gamma ^{\prime
}$ parallel to the imaginary axis in the complex $s$-plane. The real
parameter $\gamma ^{\prime }$ can be chosen arbitrarily as long as all
singularities remain to the left-hand side of $s=\gamma ^{\prime }$. After a
simple partial fraction decomposition of in Eq. (\ref{bromwich1}) we obtain,
\begin{equation}
\psi (x,t)=\frac{1}{2\pi i}\left[ \int\limits_{\gamma ^{\prime }-i\infty
}^{\gamma ^{\prime }+i\infty }\Phi _{+}(s)ds+\int\limits_{\gamma ^{\prime
}-i\infty }^{\gamma ^{\prime }+i\infty }\Phi _{-}(s)ds\right] ,
\label{notrans}
\end{equation}
where the integrands $\Phi_{\pm }(s)$ are defined as, 
\begin{equation}
\Phi _{\pm }(s)=\frac{i\beta }{2}\frac{e^{ipx}e^{st}}{p(p\pm k_{0})}.
\label{Integrand}
\end{equation}
In order to evaluate the integral (\ref{notrans}) we perform the change of
variable, $s^{\prime }=s+iV$, which allows us to identify in Eq. (\ref
{notrans}) the integral representation of the Moshinsky $M$-functions \cite
{Moshy,PRA97}
\begin{equation}
M(x,\pm k_{0},t)=\frac{1}{2\pi i}\int\limits_{\gamma -i\infty }^{\gamma
+i\infty }\frac{i\beta }{2}\frac{e^{i\sqrt{\beta is^{\prime }}x}e^{s^{\prime
}t}}{\sqrt{\beta is^{\prime }}(\sqrt{\beta is^{\prime }}\pm k_{0})}%
ds^{\prime },
\label{mosh1}
\end{equation}
and obtain the solution for the case of $\omega_0>V$, given
by Eq. (\ref{propagation}). The solution for the case of $\omega_0<V$,
given by Eq. (\ref{evanescent}), is readily obtained along the same 
lines as Eq. (\ref{propagation}).

\subsection{\label{app:subsec}Asymptotic behavior of the solutions}

In this subsection we consider the long time behavior of the solutions 
given by Eqs. (\ref{propagation}) and (\ref{evanescent}). 
This can be analyzed using the series expansion in powers of the argument
$y_q$ (Eq. (\ref{argument})) of the $M$-functions (Eq. (\ref{Mosh2})) \cite{PRA97}. 
For large values of the argument $y_q$ {\it i.e.} $|y_q|\gg 1$,
the $M$-functions have the following series representation (see Ref. 
\cite{PRA97}):
\begin{subequations}
\begin{eqnarray}
M(y_{q})\approx \frac{1}{2}e^{imx^{2}/2\hbar t}[\frac{1}{\pi ^{1/2}y_{q}}-%
\frac{1}{\pi ^{1/2}y_{q}^{3}}+...],
\label{m1}
\end{eqnarray}
\end{subequations}
provided that the phase, $\phi_q$, of $y_{q}\equiv |y_{q}|\exp(\phi_q)$ 
lies in the interval $-\pi /2<\phi_q<\pi /2$.

For the case $\pi /2<\phi_q<3\pi /2$, instead of Eq. (\ref{m1}) we have,
\begin{subequations}
\begin{eqnarray}
M(y_{q})\approx \frac{1}{2}e^{imx^{2}/2\hbar t}[2e^{y_{q}^{2}}+\frac{1}{\pi
^{1/2}y_{q}}- \nonumber\\
\frac{1}{\pi ^{1/2}y_{q}^{3}}+...].
\label{m2}
\end{eqnarray}
\end{subequations}

In general, the long time regime ($t\rightarrow \infty )$ corresponds to large 
values of the argument $y_q$. For the particular values of $q=\pm iq_0$, 
we have that $y_q \rightarrow -e^{-i\pi /4}(t/\beta )^{1/2}q$. Note that
this expression depends on the choice of $q_{0}$, and by inspection, we can
see that for the cases $q=iq_0$ and $q=-iq_0$ the arguments are
$\phi_{iq_0}=-5\pi /4$ and $\phi_{-iq_0}=-\pi /4$, respectively. Therefore 
according to Eqs. (\ref{m1}) and (\ref{m2}), one obtains that in the limit
$t\rightarrow \infty $, $M(y_{-iq_0})\rightarrow 0$ and $M(y_{iq_0})\rightarrow
e^{-q_0 x}e^{i(V-\omega_0) t}$. Substituting this values in Eq. (\ref
{evanescent}) yields the asymptotic solution given by Eq. (\ref{asinto}).
By performing a similar analysis for the case of wave evolution for $\omega_0>V$ 
given by Eq. (\ref{propagation}), one finds that the asymptotic
behavior as $t\rightarrow\infty$ is given by Eq. (\ref{asintok}).

\end{document}